# Pulse-assisted magnetization switching in magnetic nanowires at picosecond and nanosecond timescales with low energy

Furkan Şahbaz, Mehmet C. Onbaşlı*

Koç University, Department of Electrical and Electronics Engineering, Sarıyer, 34450 Istanbul

*corresponding author: monbasli@ku.edu.tr

Detailed understanding of spin dynamics in magnetic nanomaterials is necessary for developing ultrafast, low-energy and high-density spintronic logic and memory. Here, we develop micromagnetic models and analytical solutions to elucidate the effect of increasing damping and uniaxial anisotropy on magnetic field pulse-assisted switching time, energy and field requirements of nanowires with perpendicular magnetic anisotropy and yttrium iron garnet-like spin transport properties. A nanowire is initially magnetized using an external magnetic field pulse (write) and self-relaxation. Next, magnetic moments exhibit deterministic switching upon receiving 2.5 ns-long external magnetic pulses in both vertical polarities. Favorable damping ($\alpha \sim 0.1$-$0.5$) and anisotropy energies ($10^4$-$10^5$ J·m$^{-3}$) allow for as low as picosecond magnetization switching times. Magnetization reversal with fields below coercivity was observed using spin precession instabilities. A competition or a *nanomagnetic trilemma* arises among the switching rate, energy cost and external field required. Developing magnetic nanowires with optimized damping and effective anisotropy could reduce the switching energy barrier down to $3163 \times k_B T$ at room temperature. Thus, pulse-assisted picosecond and low energy switching in nanomagnets could enable ultrafast nanomagnetic logic and cellular automata.

# I.     Introduction

An in-depth understanding of spin relaxation in magnetic nanostructures is necessary to develop highly efficient and ultrafast switching methods. The interplay between external magnetic field and magnetic material properties remains to be understood at sub-100 nanosecond and nanometer length scales for fundamental studies of spin-spin, spin-electric field and spin-magnetic field interactions for developing future spintronic devices. The effects of external field amplitude [1], frequency [2] and polarization [3] on the spin relaxation of nanomagnetic media have been investigated. Previous studies indicate that switching field decreases when the polarization direction and frequency of the circularly polarized microwave field matches that of the ferromagnetic resonance of the nanomagnet [3]. Applying a microwave magnetic field with optimal frequency at or near ferromagnetic resonance reduces the coercive field by helping overcome the effective energy barrier of the domain nucleation [1]. The coercivity reduction is larger than the microwave field magnitude $H_{rf}$ at a certain frequency and input power [2].

While microwave, laser or heat-assisted switching effects that determine switching energy have been investigated [4,5], the intrinsic magnetic material property dependence of spin relaxation has not been studied extensively. Previous spin relaxation studies include nanodot models [6], permalloy rectangle models [7], nanowire models [8], and ferromagnetic nanoparticles [9]. The key magnetic properties that affect spin relaxation include Gilbert damping constant, saturation magnetic moment, exchange stiffness, anisotropy, dimensions and aspect ratios. In this study, we use analytical and numerical micromagnetic models to quantify the regimes under which increasing damping, uniaxial anisotropy and external pulse field can switch magnetism in sub-100 nm nanowires. The results of these analyses prompted us to propose an external magnetic field pulse-assisted magnetization reversal mechanism that could enable sub-coercivity and sub-nanosecond nonvolatile switching with low energy (a few thousand $k_BT$ per bit at room temperature).

Previous studies show that damping plays a key role in magnetization dynamics of nanostructures [10,11]. Ref. [10] inspected the effect of damping on reversal time without anisotropy and showed that magnetization reversal time can increase (decrease) with increasing (decreasing) damping constant. Since realistic materials have nonzero intrinsic magnetic anisotropy, magnetization reversal models must include anisotropy. A generalized analysis of magnetization reversal [12] highlights the significance of demagnetization factors and anisotropy parameters. In nanostructures, magnetoelastic [13], magnetocrystalline [14], off-stoichiometry [15], growth-induced anisotropy [16, 17] and surface-induced anisotropy (especially for large surface area-to-volume ratio nanostructures) can be modeled with an overall uniaxial anisotropy term, which alters switching times significantly [18,19]. One could engineer these terms to achieve perpendicular magnetic anisotropy (PMA) preferred in high-density memory [20]. Large perpendicular anisotropy increases the effective field and causes precession-driven magnetization dynamics with high precession frequencies [21].

We present the results of our analytical spin relaxation model and numerical methods in Section II. In section III, we present Gilbert damping constant and uniaxial anisotropy dependence of spin relaxation and magnetization reversal time. In section IV, magnetization switching time and energy are modeled functions of external magnetic field pulse intensity and width.

## II. Numerical Modeling and Analytical Solutions of Spin Relaxation

### 1. Numerical model details

Numerical models were developed to understand the magnetic relaxation and reversal in nanowires. We used Object-oriented Micromagnetic Framework (OOMMF) to obtain magnetic nanowire hysteresis loops and spin relaxation dynamics as function of damping ($\alpha$) and uniaxial anisotropy constant ($K_u$). A rectangular 20×100×10 nm$^3$ Y$_3$Fe$_5$O$_{12}$ (YIG) nanowire (width, length, thickness) was used for all models in this study. The time evolution of $m_x$, $m_y$ and $m_z$ vectors were calculated with minimum temporal step sizes of 2.14 fs. These nanowire dimensions were chosen to elucidate the effect of damping ($\alpha$=10$^{-4}$–10) and uniaxial anisotropy ($K_u$=10$^3$–10$^6$ J·m$^{-3}$) in the near single domain regime. These dimensions are experimentally feasible with state-of-the-art fabrication techniques [22-26]. We chose YIG (exchange stiffness $A_{ex}$=3.65±0.38 pJ·m$^{-1}$, saturation magnetization $M_s$=140 kA·m$^{-1}$) due to its very low and tunable damping [27-29] and due to its lower exchange stiffness compared with permalloy (13 pJ·m$^{-1}$) [7], cobalt-platinum multilayers as well as Heusler alloys (15 pJ·m$^{-1}$) [30, 31]. We focus on magnetic insulators (MI) like YIG over metals due to their reduced Joule dissipation, lower damping and lower exchange stiffness. Lower exchange stiffness and exchange energy in MI allow write energy per bit could be lower for MI than for metals. The magnetic field pulses applied on nanowire were chosen to be 2 ns wide, as Si CMOS can operate at similar periods for read/write memory pulses.

The switching models were prepared in three steps:

*1) Self-relaxation (0-15ns).* First, nanowires with different uniaxial anisotropy constants but identical geometries were allowed to equilibrate into minimum energy states in absence of external magnetic field or initial magnetization. The magnetization profiles after self-relaxation for nanowires with increasing uniaxial anisotropy constants were calculated and are shown in Fig. 1. When uniaxial anisotropy constant is low (10$^3$ J·m$^{-3}$), shape anisotropy renders the nanowire an in-plane easy axis material. When uniaxial anisotropy is large enough to overcome shape anisotropy, the nanowire becomes PMA. For lower field and lower energy switching, perpendicular magnetic anisotropy (PMA) in the nanowires is desired.

*2) Initialization (15-45ns).* In the second step, we applied an external magnetic field pulse of 2.2 Tesla for initialization of magnetic moments along +z axis.

*3) Deterministic switching (45-100 ns).* In this third and final step, 2 ns-wide and apart field pulses were applied to investigate the effect of anisotropy and damping on switching time and energy of the nanowire.

## 2. Analytical model results

The Landau-Lifshitz-Gilbert equation (Supplementary Materials Part 1) captures the time evolution in nanomagnets. Its analytical solutions yield three general cases based on the main parameters, which determine switching likelihood and time constants: precession-driven, damping-driven and effective field-driven regimes. In the precession-driven regime ($\alpha \ll 1$), magnetization reversal cannot settle since the nanomagnet undergoes precession indefinitely:

$$\mathbf{M}(\mathbf{r}, t) = M_s e^{-\kappa t}(-\hat{\mathbf{y}}\sin(\bar{\gamma}H_{eff}t) + \hat{\mathbf{x}}\cos(\bar{\gamma}H_{eff}t)) \tag{10}$$

In the damping-driven regime ($\alpha \gg 1$), the spins dissipate the injected pulse energy before triggering any magnetization reversal:

$$\frac{\partial \mathbf{m}}{\partial t} \approx \left(-|\bar{\gamma}|\alpha \, \mathbf{m} \times (\mathbf{m} \times \mathbf{H_{eff}})\right) \tag{14}$$

$$\mathbf{M}(\mathbf{r}, t) = e^{-|\bar{\gamma}|\alpha \Delta^* t}(\hat{\mathbf{x}}M_{x0} + \hat{\mathbf{y}}M_{y0}) + \hat{\mathbf{z}}M_{z0} \tag{15}$$

The effective field-driven case contains multiple in and out-of-plane anisotropy field terms that assist magnetization reversal. Here, the reversal time constant is determined by the external field, demagnetizing field and damping constant. Overall, an optimal window of damping and uniaxial anisotropy constants were found to enable deterministic magnetization reversal in picoseconds.

### III. Uniaxial anisotropy and Gilbert Damping dependence of magnetization reversal

In this section, we investigate the effect of $K_u$ on the self-relaxation and pulse-assisted switching. Fig. 1(a)-(d) show the time evolution for magnetic moments of the nanowires with $K_u = 10^3$, $10^4$, $10^5$ and $10^6$ J·m$^{-3}$, respectively, during self-relaxation (no external field applied: 0-15 ns) and during applied external magnetic field pulse (15-45 ns) and after the pulse is applied (45-100 ns). The nanowires were first set to an infinitesimally small magnetic moment and they were allowed to relax their magnetic moments in absence of external magnetic field until 15 ns. This initialization numerically demonstrates the easy axis for each case before applying the magnetic fields. The magnetic moment of the nanowire in Fig. 1(a) with $K_u = 10^3$ J·m$^{-3}$ self-relaxes towards +y direction, which indicates that its magnetic easy axis is along the long axis (y) of the structure and that $K_u < K_{shape}$. For Fig. 1(b), the structure relaxes to –z direction, indicating that uniaxial anisotropy now overcomes shape and renders the nanowire PMA. In Fig. 1(c), although $K_u = 10^5$ J·m$^{-3}$ > $K_{shape}$, the nanowire cannot relax to a vertical direction since the spins form a transient domain wall (Spin profiles in Supplementary Figure S1). When an additional external pulse was applied, the multi-domain structure overcomes the domain wall energy barrier, aligns and stabilizes along +z direction. In Fig. 1(d), the structure is clearly PMA and it relaxes to –z direction in 40 ps. Increasing uniaxial anisotropy energy from $10^3$ to $10^6$ J·m$^{-3}$ changes self-relaxation times from 2 ns (in-plane) down to 4 ns (PMA, single domain), 2 ns (PMA but two transient domains) to 40 ps (PMA, single domain), respectively.

When an external magnetic field pulse along +z axis has been applied for initialization, in each case except Fig. 1(d), the magnetic moment aligns with the external field first. Since the structure in Fig. 1(a) has in-plane easy axis, it cannot retain its moment along +z and it relaxes to surface plane. Since the structure in Fig. 1(b) is PMA, it switches to +z and retains its remanent state ($H_{external}$ = 2.2 T > $H_{sat}$ ~ $2K_u/M_s$ = 0.143 T). The applied field on the nanowire in Fig. 1(c) helps overcome the domain wall energy and helps align the domains along +z axis as the saturation field for this structure ($H_{sat}$ ~ $2K_u/M_s$ = 1.43 T) is less than the applied pulse intensity. Since the structure is intrinsically PMA, the structure retains its magnetic moment along +z. In Fig. 1(d), since the calculated $H_{sat}$ is about 14.3 T, the structure is not magnetically saturated and does not switch although it is PMA. The nanowire size determines the shape anisotropy and the minimum uniaxial anisotropy energy needed for PMA. When PMA is achieved with sufficiently large $K_u$, increasing $K_{eff}$ reduces the self-relaxation time down to sub-100 ps ranges although increasing $K_u$ to as high as $10^6$ J·m$^{-3}$ increases the saturation field beyond feasible magnetic field intensities.

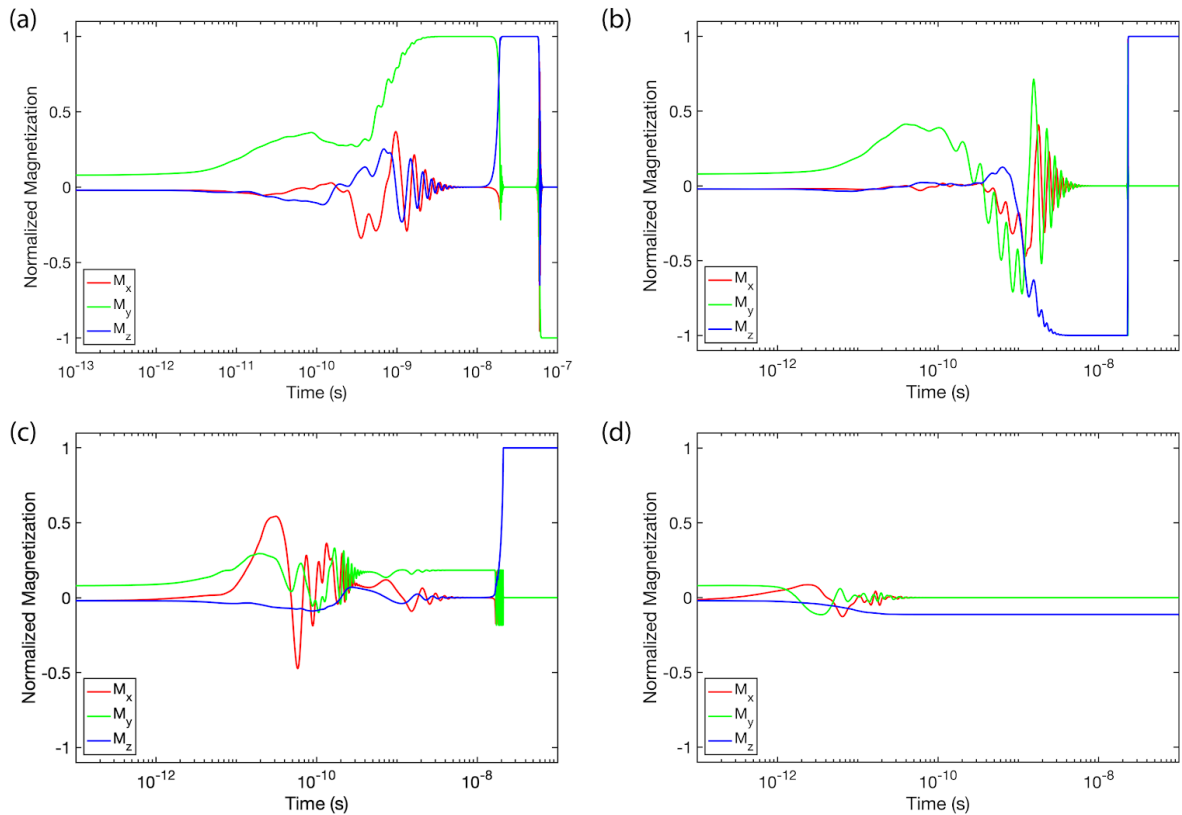

**FIG. 1. Magnetic initialization steps of the nanowires with α = 0.1 for (a-d) $K_u$ = $10^3$, $10^4$, $10^5$, and $10^6$ J·m$^{-3}$, respectively.**

Deterministic switching has been shown by applying six consecutive positive and negative switching pulses with 2200 mT intensity and 2 ns width each. Fig. 2 shows the corresponding switching time as a function of $K_u$ and $\alpha$. The colored regions indicate deterministic switching with times corresponding to their color codes. The gray regions indicate no deterministic switching. Switching time is defined as the time it takes for transitioning from $m_z = -1$ to $+1$ (or vice versa) upon receiving an external magnetic field pulse. For $\alpha > 0.1$, as uniaxial anisotropy increases, the total effective field $H_{eff}$ of nanowire increases and the switching time increases due to longer precession. As $\alpha$ increases, switching time decreases as the damping term starts balancing the precession term in the LLG equation. In the ideal case of no damping, the spins would have precessed indefinitely at $\omega = \gamma H_{eff}$ without aligning with the applied external magnetic field. With finite or increasing damping term, the precession energy is absorbed and the spins equilibrate faster. Deterministic switching was not observed for materials with low damping ($\alpha < 10^{-1}$) as precession prevented switching.

For $\alpha > 10^{-1}$, relaxation timescales are reduced to below mostly 400 ps. When uniaxial anisotropy energy exceeds $5 \times 10^4$ J·m$^{-3}$ until $1.5 \times 10^5$ J·m$^{-3}$, the nanowire starts forming domain walls, which prevents from or delays reaching steady state reversal. As the uniaxial anisotropy energy increases towards $5 \times 10^4$ J·m$^{-3}$, the domain wall width $\delta_{DW} = 2\sqrt{(A/K_u)}$, decreases to 17 nm which is below the nanowire width (20 nm). With higher uniaxial anisotropy energies, domain wall width decreases and domains form within the nanowire. For $K_u > 1.5 \times 10^5$ J·m$^{-3}$, saturation field exceeds the applied external field pulse (2200 mT) and the nanowires are not fully saturated. Therefore, for multi-domain grains or nanostructures, anisotropy energy must be large enough to achieve PMA and $K_u$ should be sufficiently small such that realistic external field pulse intensities could reverse magnetic orientation.

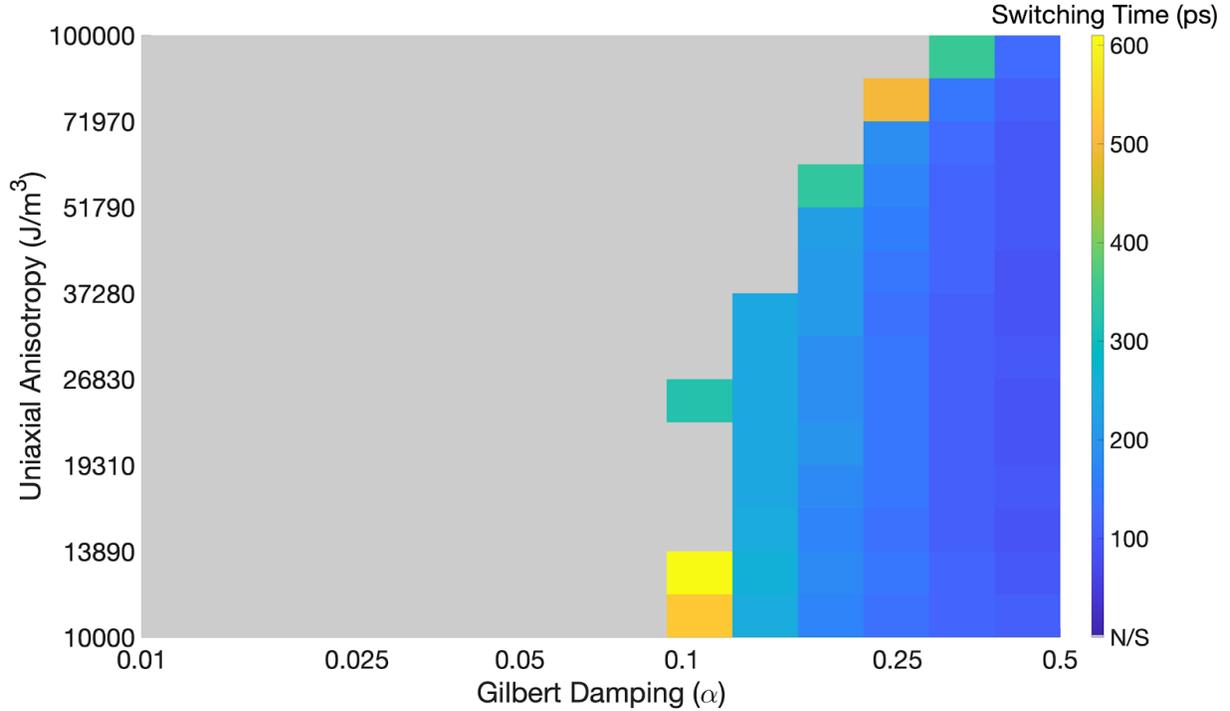

**FIG. 2. Gilbert damping and uniaxial anisotropy constant dependence of switching time ($H_{ext}$=2200 mT).**

### IV. Dependence of switching energy and rate on pulse width and intensity

Fig. 3(a) shows the switching energy of the nanowire in units of $k_BT$ (T = 300 K) for pulse widths between 50 and 3000 ps and external magnetic field intensities between 1000 and 10000 mT. The energy barriers were calculated in micromagnetic models based on the energy magnitudes the nanowire overcomes after applying the external magnetic field pulse. In these calculations, the energy difference accounts for the Zeeman, demagnetizing, exchange and uniaxial anisotropy energies. In the Hamiltonian (Suppl. eqn. 3-6), the time evolution of the energy is driven mainly by the changes in the Zeeman energy due to reorientation of the nanowire spins upon applying field pulse and the demagnetizing field of the geometry. In this figure, the gray regions show no switching (N/S) and the other regions have switching energies corresponding to their color codes. The figure indicates that the switching energy of the nanowire could be lowered from over 12000

$k_BT$ to 3163 $k_BT$ by tuning the applied field pulse. This effect indicates low Zeeman energy (due to its reduced volume), low uniaxial anisotropy and low exchange stiffness (of YIG) make magnetization reversal energetically favorable. Thus, decreasing the field intensity decreases the switching energy.

The hysteresis loop calculated for the $m_z$ component indicate that nanowires have a coercivity of 1950 mT with PMA (Supplementary Figure S4). The nanowire switches its magnetic orientation even below this coercivity with the external field pulse. Decreasing the pulse width helps reduce switching energy until applied field pulses of 4394 mT, since it reduces the average Zeeman energy injected into the nanowire. Therefore, Fig. 3(a) indicates that pulse-assisted and sub-coercivity switching with lower energy costs can be achieved for magnetic nanowires. Fig. 3(b) shows nanowire switching time for the same pulse widths and intensities used in Fig. 3(a). The switching time was calculated based on the time the nanowire takes for a complete steady-state reversal of its vertical magnetization. The fastest magnetization reversal occurs at 0.150 ns for 10000 mT and 368.4 ps pulse width. The large field intensity and short pulses enable fast magnetization reversal and minimal time spent in transient precession motion. Decreasing the field intensity increases the switching time. While the nanowire switches faster for fields above its coercive field (1950 mT), one could achieve complete magnetization reversal with sub-coercivity pulses with longer switching times. Switching with sub-coercivity pulses relies on the dynamic instabilities of the magnetic moment and most sub-coercivity switching cases in Fig. 3(b) have extended switching times. Low field intensities cause precession for extended periods (damping-dominated regime), thus preventing or significantly delaying reversal. Decreasing the pulse width helps reduce switching time as it reduces the interaction time between the magnetic moment and the field. Pulse-assisted and sub-coercivity switching could be achieved for magnetic nanowires if longer transient reversal times are allowed. Thus, a trade-off between optimal switching energy/ time and pulse width/field intensity could be established.

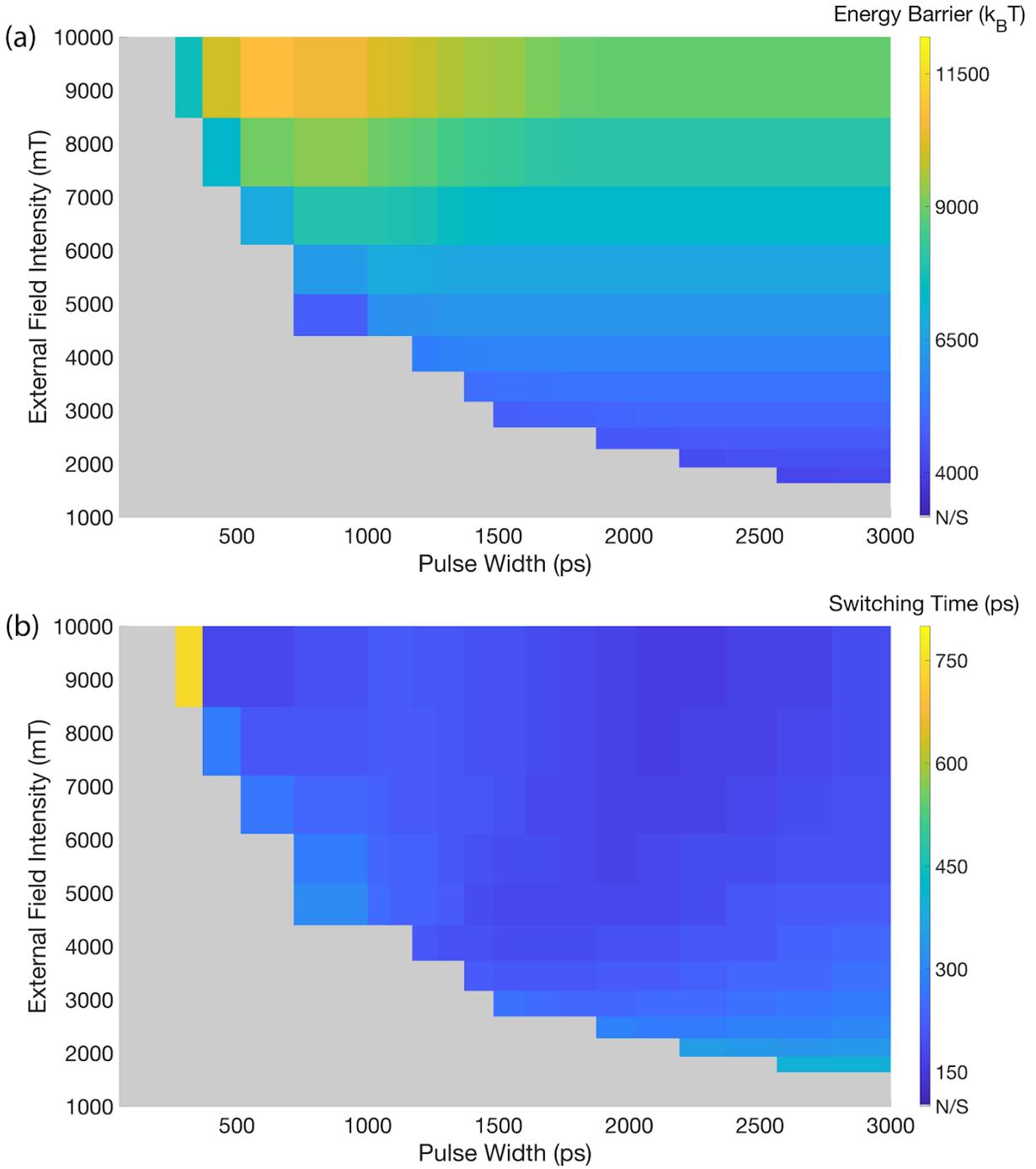

**FIG. 3.** Pulse width and intensity dependence of (a) switching energy (units of $k_B T$ at T = 300K) and (b) switching time for the nanowire with $K_u = 10^4$ J·m$^{-3}$ and $\alpha = 0.01$.

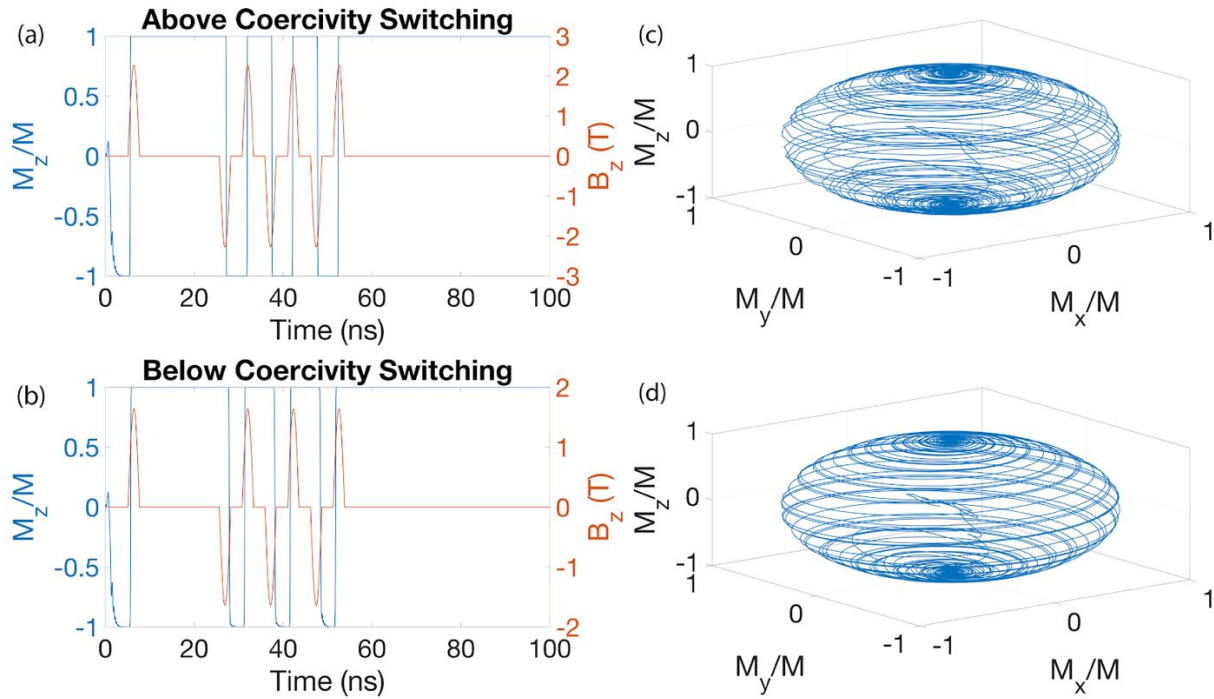

**FIG. 4. Deterministic switching for cases with short and longer switching times.** External field dependence of relaxation rate for (a) and (b) with external field intensity: 2276 mT, pulse width: 2564.3 ps (faster) ($\alpha = 0.01$, $K_u = 10^4$ J·m$^{-3}$), and for (c and d) external field intensity: 1638 mT, pulse width: 2564.3 ps (sub-coercivity, slower) ($\alpha = 0.01$, $K_u = 10^4$ J·m$^{-3}$),

Based on the deterministic switching results shown on Fig. 3, we investigate further two cases from Fig. 3(a) and (b): (pulse intensity, pulse width) = (2276 mT, 2564.3 ps) and (1638 mT, 2564.3 ps). For this condition, the calculated hysteresis loops indicate that the saturation field is 1950 mT (Suppl. Fig. S4). These two cases were chosen to investigate the deterministic switching dynamics for above and below-coercivity switching, respectively. The switching dynamics of the first and second cases are shown on Fig. 4(a,b) and Fig. 4(c,d), respectively. Fig. 4(a,b) demonstrate shorter switching times due to the higher external field intensity on the nanowire. Fig. 4(c,d) show deterministic magnetization switching at sub-coercivity external fields (below 1950 mT). As shown on Fig. 3, sub-coercivity deterministic switching requires the pulse duration to be greater than a minimum threshold (i.e. 2192 ps for 1931 mT). This threshold depends on both the extrinsic factors (external field intensity, nanowire dimensions) and intrinsic factors ($K_u$, $\alpha$, $A_{ex}$ and $M_s$). Magnetic reversal is delayed due to precession-driven switching dynamics. These results show that deterministic sub-coercivity switching in magnetic nanowires is feasible and allows for reduced switching fields with lower energy barrier materials and geometries.

## V. Conclusions

The temporal and spatial evolution of magnetization switching in nanowires were investigated as functions of pulse width, pulse intensity, uniaxial anisotropy constant and damping. Damping, precession and effective field-driven regimes have been identified in the analytical models of magnetization reversal in nanowires. These simulations and models indicate that the magnetization states of these magnetic nanowires could be reversed under external pulses with sufficient pulse intensity and width for optimal damping ($\alpha > 0.1$) and uniaxial anisotropy ($K_u < 10^5$ J·m$^{-3}$). In high aspect ratio nanowires (in plane x:y = 100:20), sufficiently high uniaxial anisotropy constants $K_u$ (at least $10^4$ J·m$^{-3}$) are needed to obtain perpendicular magnetic anisotropy by overcoming shape anisotropy. When $K_u$ becomes too high ($\geq 10^5$ J·m$^{-3}$), the effective anisotropy of nanowire increases beyond feasible magnetic field pulse intensities (2.2 T or less). For optimal damping, anisotropy and pulse properties ($\alpha \in [0.1, 0.5]$, $K_u \in [10^4, 10^5$ J·m$^{-3}$), 0.5 to 3ns-wide pulses), the nanowires could switch with picosecond timescales and low energy consumption per bit as low as $3.163 \times 10^3$ $k_B T$ at T = 300 K. In all switching cases, PMA is necessary for deterministic pulse-assisted magnetization reversal and dense memory bits. The effective field provides the energy barrier needed for both stable memory and low-power logic functionalities.

Two key outcomes emerge from this study: First is the observation of a nanomagnetic switching trilemma or the competition between nanowire (i) switching rate, (ii) energy cost of switching per bit and (iii) external field required for switching. In this trilemma, high switching rate requires either high external field or high switching energy. Lower switching energy requires an optimal external magnetic field intensity and pulse width per bit. Minimizing the external magnetic field and reducing switching time requires optimal pulse width at the cost of increasing energy per bit. This trilemma originates from a more general competition between the energy-delay product between the external magnetic field and the damping-driven magnetization reversal (Suppl. Fig. S3). The second key outcome is sub-coercivity switching observed under appropriate uniaxial anisotropy (i.e. $K_u=10^4 J \cdot m^{-3}$), damping ($\alpha = 0.1$) and pulse properties (1638<intensity <1931 mT, 2564<pulse width<3000 ps). The results shown in Fig. 3(b) demonstrate that the nanowire geometry is appropriate for deterministic switching with external magnetic field pulses with peak intensities below the coercive field. Engineering the nanowire geometry and perpendicular effective anisotropy ($K_{eff}$) reduces the switching time to sub-150 ps ranges with sub-coercivity switching. Engineering the energy barrier through nanostructure geometry optimization and operating at the optimal point of the nanomagnetic trilemma could pave the way for efficient memory, logic and cellular automata at high bit rates at room temperature.

**Acknowledgements**

F. Ş. acknowledges Koç University Scholarship. M.C.O. acknowledges BAGEP 2017 Award, TÜBA-GEBİP Award by Turkish Academy of Sciences and TUBITAK Grant No. 117F416.

**Supplementary Information for "Pulse-assisted magnetization switching in rectangular magnetic nanowires at picosecond and nanosecond timescales with low energy"**

Furkan Şahbaz, Mehmet C. Onbaşlı*

Koç University, Department of Electrical and Electronics Engineering, Sarıyer, 34450 Istanbul

*corresponding author: monbasli@ku.edu.tr

In this supplementary information, we provide the additional numerical modelling results and theoretical analyses in two sections: (1) Numerical Modelling and (2) Theoretical Analysis. The appendix to these sections consists of our OOMMF source codes (mif files).

1. Numerical Modelling

*Transient domain wall formation:* In Fig. 1(c) of the main manuscript, we observed that the spins cannot relax to a vertical direction despite uniaxial energy exceeding shape anisotropy energy ($K_u > K_{shape}$). We attribute the origin of this effect in the main manuscript to a transient domain wall (DW) formation. In Supplementary Figure S1, we present the calculated spin profile for $K_u = 10^5$ J·m$^{-3}$. Fig. S1 shows a DW, which traverses the nanowire along its short axis.

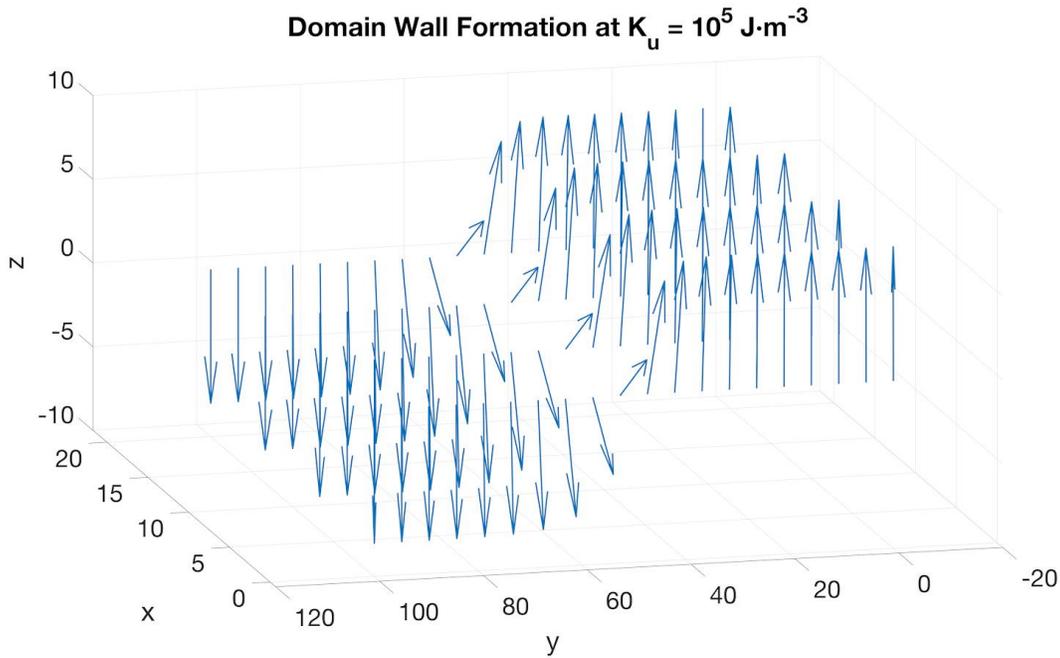

**Supplementary Figure S1. For $K_u = 10^5$ J·m$^{-3}$, a domain wall prevents relaxation to the vertical axis (z) (the units are in nm for all three axes).**

*Nondeterministic reversal of nanomagnets over a wider window of uniaxial anisotropy and Gilbert damping values:* Fig. 2 of the main manuscript includes results on Gilbert damping ($\alpha$ = 0.01-0.5) and uniaxial anisotropy ($K_u$ = $10^4$-$10^5$ J·m$^{-3}$) dependence of relaxation rate. We calculated the relaxation rates for the range of $\alpha$ = $10^{-4}$-$10^{-1}$ and $K_u$ = $10^3$-$10^6$ J·m$^{-3}$. We observed that switching is not deterministic or cannot happen at all under these conditions. In Supplementary Figure 2, we indicate this result as a lack of deterministic switching case (N/S: no switching). For $K_u$ > $10^6$ J·m$^{-3}$, the applied external magnetic field is not sufficient to overcome the effective magnetic field. For $K_u$ < $10^6$ J·m$^{-3}$, damping constant is not sufficient to stabilize the spin precession during magnetization reversal.

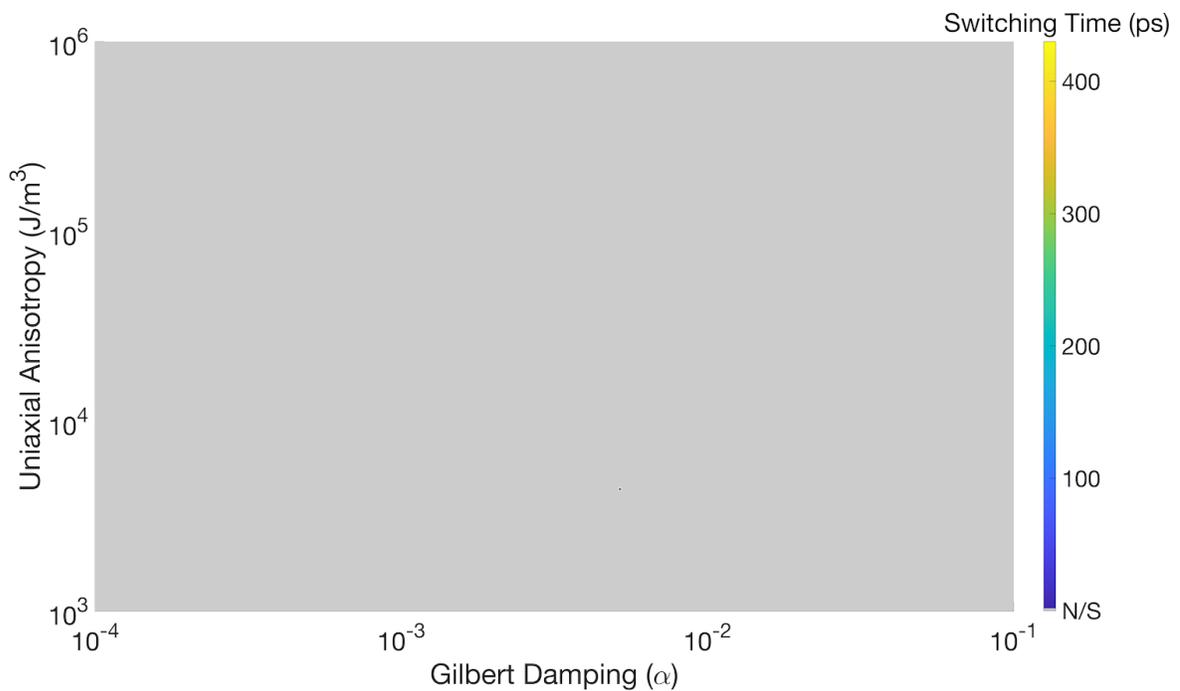

**Supplementary Figure S2. The calculated relaxation rates for the range of $\alpha$ = $10^{-4}$-$10^{-1}$ and $K_u$ = $10^3$-$10^6$ J·m$^{-3}$. For $K_u$ > 1.5 × $10^5$ J·m$^{-3}$, the saturation field exceeds the applied external field pulse and the nanostructures are not saturated. For lower anisotropy values, although magnetization reversal occurs, the damping is not sufficient to end the precession motion for deterministic switching.**

*Energy-delay product and the nanomagnetic trilemma:* In the main text, we mentioned the competition between (i) switching rate, (ii) energy cost of switching per bit and (iii) external field required for switching. We named this competition the nanomagnetic trilemma since this effect originates from a more general competition between the energy-delay product between the external magnetic field intensity and the internal precession/damping-driven reversal mechanisms of magnetic nanostructures. In Supplementary Fig S3, we provide the calculated energy-delay product (units of fJ·ps) for different pulse widths and external field intensities.

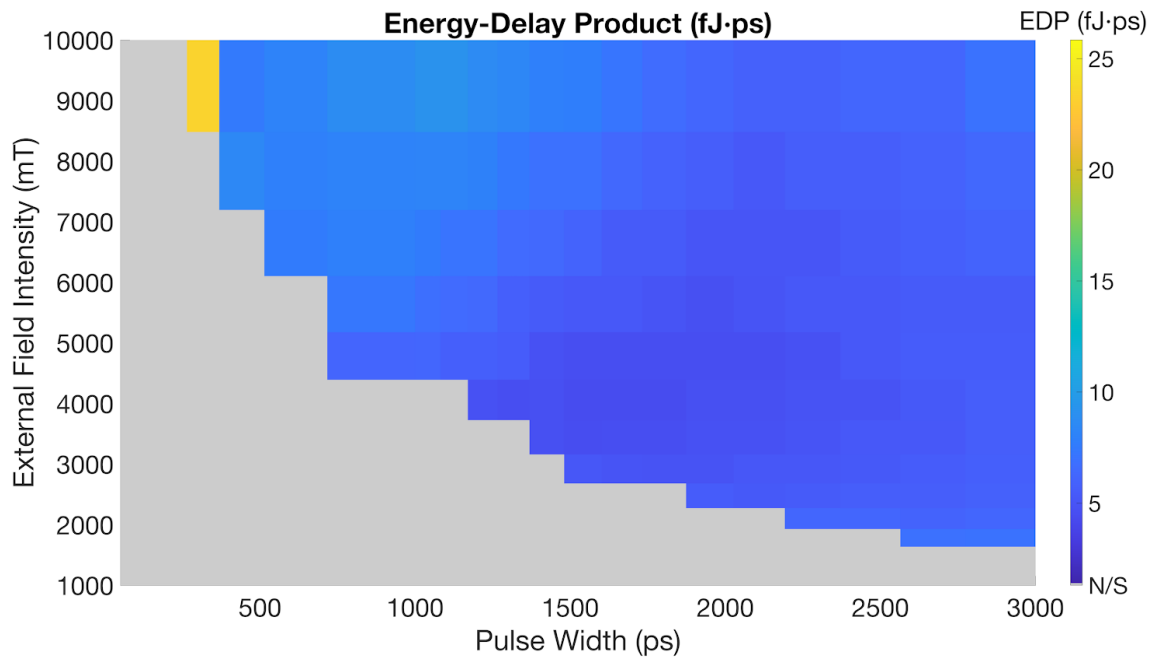

**Supplementary Figure S3. Energy-delay product (EDP, fJ·ps) for the external field-driven magnetization reversal and the internal precession/damping-driven reversal mechanisms of magnetic nanostructures.**

*Energy-delay product and the nanomagnetic trilemma:* In the main text, we mentioned that the calculated hysteresis loops indicate that the saturation field is 1950 mT. In Supplementary Fig. S4, we present the calculated hysteresis loop for the normalized vertical magnetic moment component $m_z$ as a function of applied magnetic field $H_z$.

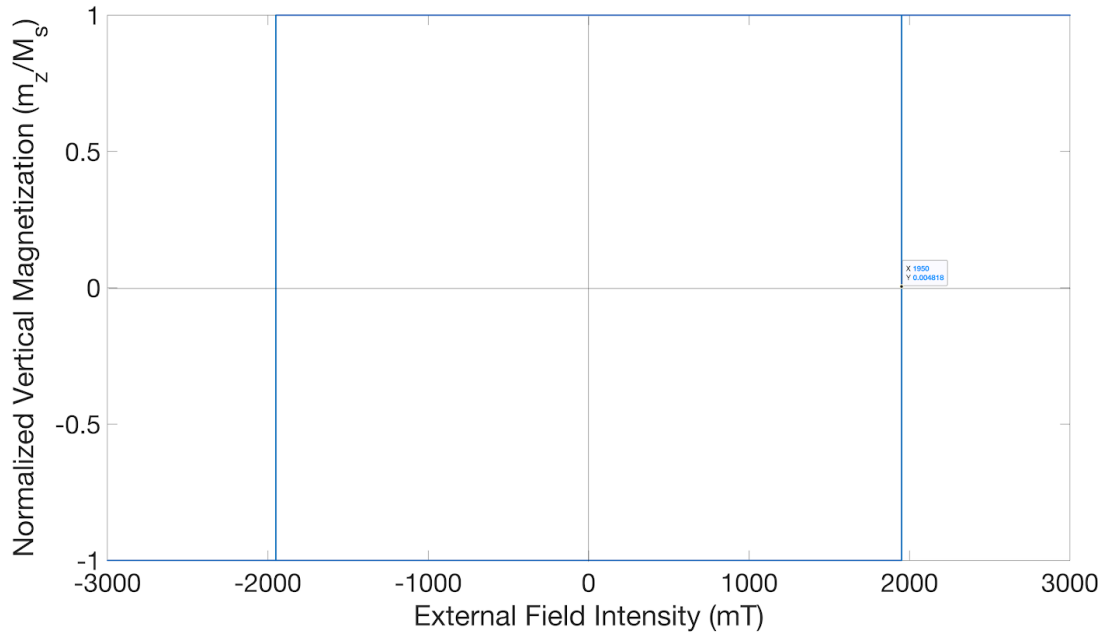

**Supplementary Figure S4. Simulated magnetic hysteresis loop of the nanowires when an external magnetic field is applied on the nanowire along vertical axis (z). The hysteresis loop shows the normalized vertical magnetization component ($m_z$), a perpendicular magnetic easy axis with a vertical remanent state and coercive and saturation field of $H_c = H_{sat} = 1950$ mT.**

1. **Theoretical Analysis**

Dynamic evolution of spin vector components in a magnetic material is described using the Landau-Lifshitz-Gilbert (LLG) shown in equations (1) and (2). Here, **m** is the normalized magnetization vector with $\mathbf{m} = \frac{\mathbf{M}(x,y,z,t)}{M_s}$, where $\mathbf{M}(x,y,z,t)$ is the magnetization profile throughout the magnetic nanostructure and $M_s$ is the saturation magnetic moment of the material.

$$\frac{\partial \mathbf{m}}{\partial t} = -\gamma \mathbf{m} \times \mathbf{H}_{\text{eff}} + \alpha \mathbf{m} \times \frac{\partial \mathbf{m}}{\partial t} \tag{1}$$

or

$$\frac{\partial \mathbf{M}}{\partial t} = -|\bar{\gamma}|\mathbf{M} \times \mathbf{H}_{\text{eff}} - \frac{|\bar{\gamma}|\alpha}{M_s}\mathbf{M} \times (\mathbf{M} \times \mathbf{H}_{\text{eff}}) \tag{2}$$

where $\bar{\gamma}$: Landau − Lifshitz gyromagnetic ratio, α: damping coefficient

In this equation, time evolution of magnetic moment vectors sampled within a rectangular grid of 5 nm size is calculated over the rectangular magnetic nanostructure presented above. $\mathbf{H}_{\text{eff}}$ is the effective magnetic field vector. As the total energy of the nanostructure is minimized along perpendicular axis, Equation (3) describes the magnetic anisotropy and perpendicular easy axis of the magnetic nanostructure:

$$\mathbf{H}_{\text{eff}} = \mathbf{H}_{\text{external}} + \mathbf{H}_{\text{demag}} + \mathbf{H}_{\text{exchange}} + \mathbf{H}_{\text{uniaxial}} \tag{3}$$

$$\mathbf{H}_{\text{external}} = \hat{\mathbf{z}} H_0 \sin(\omega t) \tag{4}$$

$$\mathbf{H}_{\text{demag}} = \hat{\mathbf{y}} H_1 \tag{5}$$

$$\mathbf{H}_{\text{exchange}} = \hat{\mathbf{z}} \frac{2A_0}{\mu_0 M_s} \nabla^2 m \tag{6}$$

$$\mathbf{H}_{\text{uniaxial}} = \hat{\mathbf{z}}(K_1 \sin^2\theta + K_2 \sin^2\theta \sin^2\alpha) \approx \hat{\mathbf{z}} K_1 \sin^2\theta \tag{6}$$

The first term of the right hand side of Equation (2) is known as the precession term that drives oscillations within nanowires under magnetic fields. The second term, also called the damping term, drives the alignment rate of the magnetic moment with the external magnetic field. This dissipative term is one of the energy loss channels in the relaxation process. The "Hamiltonian" for the LLG equation or the effective field is defined in Equation 3. This equation includes the contributions from external magnetic field, demagnetization field (shape anisotropy term), exchange field term and uniaxial anisotropy field. The external field, $H_{\text{external}}$, is applied perpendicular to the nanostructure surface along z axis. The demagnetization term, $H_{demag}$, is the field, which originates due to the absence of magnetic monopoles (divergence-free magnetic flux density) and the resulting field distribution within the nanowire geometry along its long in-plane y axis. Demagnetizing field is one of the terms, which captures the effect of geometry on anisotropy and magnetization dynamics. The exchange term, $H_{exchange}$, is a field generated due to the Heisenberg exchange interaction between adjacent spins. This field can become particularly important when metallic magnetic nanowires are used (permalloy: $A_{ex}$ = 13 pJ·m$^{-1}$) [1] and Cobalt-Platinum multilayers ($A_{ex}$ = 15 pJ·m$^{-1}$) [2]) and less significant for YIG nanostructures with $A_{ex}$ = 3.65 ± 0.38 pJ·m$^{-1}$ [3-5] although the presence of exchange interaction is not essential for spin wave propagation or magnetization reversal [6]. The last term is the uniaxial anisotropy energy term, which indicates the vertical directional preference of magnetic moment during relaxation and switching. This term could originate from a variety of sources including magnetocrystalline anisotropy, magnetoelastic anisotropy for thin epitaxial nanostructures, strain doping as well as other growth-induced uniaxial anisotropy.

One can analyze magnetic relaxation and switching in nanostructures in three regimes:

(1) precession-driven (when damping term is negligible),

(2) damping-driven (α is large such that the damping term prevents magnetization reversal)

(3) effective field-driven

Considering the combined effects of material constants and the anisotropy terms, we derive and investigate these regimes in further detail below.

### I. Precession-driven magnetization dynamics

When Gilbert damping α is small such that damping term is much smaller than the precession term, magnetic relaxation and reversal is driven by precession:

$$\left| -\frac{|\bar{\gamma}|\alpha}{M_s} \mathbf{M} \times (\mathbf{M} \times \mathbf{H_{eff}}) \right| \ll |-|\bar{\gamma}|\mathbf{M} \times \mathbf{H_{eff}}| \quad (7)$$

$$\left| \frac{\alpha}{M_s} \mathbf{M} \times (\mathbf{M} \times \mathbf{H_{eff}}) \right| \ll |\mathbf{M} \times \mathbf{H_{eff}}|$$

$$\left| \frac{\alpha \mathbf{M}}{M_s} \right| |(\mathbf{M} \times \mathbf{H_{eff}})||\sin(90°)| \ll |\mathbf{M} \times \mathbf{H_{eff}}|$$

$$\alpha|m| \ll 1$$

$$\alpha \ll 1 \quad (8)$$

Here, the angle between the magnetization vector **M** and the **M**×**H**$_{eff}$ vector is 90°. Since the magnitude of the normalized magnetization vector m is always one, precession term dominates when the damping coefficient is much smaller than one. When magnetization dynamics is driven by precession, relaxation or reversal processes continue indefinitely or much longer than otherwise in absence of damping. As a result, we do not observe any magnetization reversal:

$$\frac{\partial \mathbf{M}}{\partial t} \approx -|\bar{\gamma}|\mathbf{M} \times \mathbf{H}_{eff} \qquad (9)$$

$$\mathbf{M}(\mathbf{r},t) = M_s e^{-\kappa t}(-\hat{\mathbf{y}}\sin(\bar{\gamma}H_{eff}t) + \hat{\mathbf{x}}\cos(\bar{\gamma}H_{eff}t)) \qquad (10)$$

For κ = 0 (zero damping limit), the oscillation continues indefinitely. For nonzero and small κ, the oscillations continue for extended periods with an evanescent decaying envelope. Gilbert damping parameter α is low for YIG [4,5] (3-7×10$^{-4}$), magnetostrictive spinel ferrites [7] (<3×10$^{-3}$), Heusler [8] (10$^{-3}$) or other low-damping metallic alloys such as CoFe [9] (10$^{-4}$-10$^{-3}$). A very small damping coefficient, regardless of the H$_{eff}$ magnitude, triggers the precession-driven regime. The terms present in the effective field determine the Larmor precession frequency (around few GHz for ferromagnets and potentially towards THz for antiferromagnets).

## II. Damping-driven magnetization dynamics

When Gilbert damping α is large such that the damping term is much larger than the precession term, magnetic relaxation and reversal is driven by damping:

$$\left|-|\bar{\gamma}|\mathbf{M} \times \mathbf{H}_{eff}\right| \ll \left|-\frac{|\bar{\gamma}|\alpha}{M_s}\mathbf{M} \times (\mathbf{M} \times \mathbf{H}_{eff})\right| \qquad (11)$$

$$|\mathbf{M} \times \mathbf{H}_{eff}| \ll \left|\frac{\alpha}{M_s}\mathbf{M} \times (\mathbf{M} \times \mathbf{H}_{eff})\right|$$

$$|\mathbf{M} \times \mathbf{H}_{eff}| \ll \left|\frac{\alpha}{M_s}\mathbf{M}\right| |\mathbf{M} \times \mathbf{H}_{eff}||\sin(\theta)|$$

$$1 \ll \alpha|m||\sin(\theta)|$$

$$1 \ll \alpha \qquad (12)$$

When the magnetization dynamics is driven by damping due to large damping coefficient, relaxation or reversal processes occur with very short evanescent lifetimes:

$$\frac{\partial \mathbf{M}}{\partial t} \approx -\frac{|\bar{\gamma}|\alpha}{M_s} \mathbf{M} \times (\mathbf{M} \times \mathbf{H}_{eff}) \tag{13}$$

$$\frac{\partial \mathbf{M}}{\partial t} \approx -|\bar{\gamma}|\alpha\, \mathbf{m} \times (\mathbf{M} \times \mathbf{H}_{eff})$$

$$\frac{1}{M_s}\frac{\partial \mathbf{M}}{\partial t} \approx \frac{1}{M_s}(-|\bar{\gamma}|\alpha\, \mathbf{m} \times (\mathbf{M} \times \mathbf{H}_{eff}))$$

$$\frac{\partial \mathbf{m}}{\partial t} \approx \left(-|\bar{\gamma}|\alpha\, \mathbf{m} \times (\mathbf{m} \times \mathbf{H}_{eff})\right) \tag{14}$$

Equation 14 shows that the decay rate in the damping-driven regime is driven by the gyromagnetic ratio, damping constant, effective field and the orientation of the effective field with respect to the initial magnetization orientation. For a magnetic moment initially oriented along +z, large damping decays the in-plane excitations due to the pulse and the initial magnetization along +z is retained:

$$\mathbf{M}(\mathbf{r}, t) = e^{-|\bar{\gamma}|\alpha\, \Delta^* t}(\hat{\mathbf{x}} M_{x0} + \hat{\mathbf{y}} M_{y0}) + \hat{\mathbf{z}} M_{z0} \tag{15}$$

In order to trigger damping-driven regime described by equations 11, 14 and 15, Gilbert damping constant α should be larger than 1. The effective field should also not be parallel to the initial magnetization, since this configuration would not trigger any reversal.

When a material has ultralow damping and has high effective fields as in Yttrium iron garnet nanowires, precession regime prevails and relaxation timescales could be extended indefinitely as long as damping is negligible or compensated. For ultrafast magnetization reversal, damping must not be negligible or effective field (anisotropy and external field) must not be too high.

### III.  Effective field-driven magnetization dynamics

When neither damping nor precession term dominates time-dependent magnetic relaxation, dynamic control of individual terms in the effective field determines the time evolution of magnetization reversal process. In this case, the LLG equation follows the standard form in equation 2. The vectors in the effective field term determine the switching time scales together with damping. The external field or the uniaxial anisotropy determine the timescales in the LLG equation when they are much larger with respect to the other terms in the effective field. The other in-plane terms in the effective field, such as the demagnetizing fields, are necessary to trigger magnetization reversal:

$$\frac{\partial \mathbf{M}}{\partial t} = -|\bar{\gamma}|\mathbf{M} \times (\mathbf{H}_{external} + \mathbf{H}_{demag} + \mathbf{H}_{exchange} + \mathbf{H}_{uniaxial}) - \frac{|\bar{\gamma}|\alpha}{M_s}\mathbf{M} \times (\mathbf{M} \times (\mathbf{H}_{external} + \mathbf{H}_{demag} + \mathbf{H}_{exchange} + \mathbf{H}_{uniaxial})) \quad (16)$$

$$\frac{\partial \mathbf{M}}{\partial t} \approx -|\bar{\gamma}|\mathbf{M} \times (\mathbf{H}_{external} + \mathbf{H}_{in\ plane}) - \frac{|\bar{\gamma}|\alpha}{M_s}\mathbf{M} \times (\mathbf{M} \times (\mathbf{H}_{external} + \mathbf{H}_{in\ plane})) \quad (17)$$

Here, achieving ultrashort reversal time constants depend on the damping constant and the external field intensity along the direction of the new magnetic state. Large external field and some nonzero built-in in-plane field drives a faster precession and magnetization reversal, while a sizable damping is necessary to stabilize the moments along the final orientation.

**Appendix 1. OOMMF Source Code**
**(rect_structure_field.mif and rect_structure_hysteresis.mif)**
**Appendix 2. Calculated domain wall movie for Supplementary Fig. S1 & Figure 1(c)**
**(1E5 Sample.mov)**